\documentclass[aip, apl, reprint, 10pt, twocolumn, showpacs, preprintnumbers,superscriptaddress, amsmath, amssymb]{revtex4-2}
\usepackage[pdftex]{graphicx}
\usepackage{subcaption}
\usepackage[ngerman,english]{babel}
\usepackage{natbib}
\usepackage{color}
\usepackage[separate-uncertainty=true]{siunitx}
\usepackage{calc}
\usepackage[colorlinks=true, pdfstartview=FitV, linkcolor=blue,citecolor=blue, urlcolor=blue]{hyperref}
\usepackage{float}

\definecolor{green2}{rgb}{.0, .58, 0}

\newcommand{\grad}{\ensuremath{^\circ}}

\begin{document}
\begin{abstract}
   We demonstrate the fabrication of scanning superconducting quantum interference devices (SQUIDs) on the apex of sharp quartz scanning probes -- known as SQUID-on-tip probes -- using conventional magnetron sputtering. We produce and characterize SQUID-on-tips made of both Nb and MoGe with effective diameters ranging from 50 to 80 nm, magnetic flux noise down to \SI{300}{\nano\Phi_{0}/\sqrt{\hertz}}, and operating fields as high as \SI{2.5}{\tesla}. Compared to the SQUID-on-tip fabrication techniques used until now, including thermal evaporation and collimated sputtering, this simplified method facilitates experimentation with different materials, potentially expanding the functionality and operating conditions of these sensitive nanometer-scale scanning probes.
\end{abstract}
\title{Fabrication of Nb and MoGe SQUID-on-tip probes by magnetron sputtering}
\author{G.~Romagnoli} \affiliation{Department of Physics, University of Basel, 4056     Basel, Switzerland}
\author{E.~Marchiori} \affiliation{Department of Physics, University of Basel, 4056     Basel, Switzerland}
\author{K.~Bagani} \affiliation{Department of Physics, University of Basel, 4056     Basel, Switzerland}
\author{M.~Poggio} \affiliation{Department of Physics, University of
  Basel, 4056 Basel, Switzerland} \affiliation{Swiss Nanoscience
  Institute, University of Basel, 4056 Basel, Switzerland}
\email{martino.poggio@unibas.ch}
\maketitle

Scanning superconducting quantum interference device (SQUID) microscopy (SSM) is among the most sensitive and least invasive methods for imaging subtle magnetic field patterns near a sample surface~\cite{marchiori_nanoscale_2022}. In recent years, SSM has been used to study microscopic properties of superconducting devices~\cite{embon_Pinning_2015,ceccarelli_imaging_2019}, nanomagnets~\cite{vasyukov_nanotube_2018}, magnetic oxides~\cite{persky_oxide_2018}, two-dimensional materials~\cite{uriMappingTwist2020, tschirhartImagingOrbitalFerromag2021}, quantum wells~\cite{spanton_EdgeCurrent_2014, nowackImagingCurrentsHgTe2013}, and topological states matter~\cite{uri_quantumHall_2020}. In general, in order to allow for both high spatial resolution and high sensitivity magnetic imaging, the probe must be miniaturized and brought as close to the sample as possible, while retaining high flux sensitivity. A number of strategies for realizing such SSM probes exist, including techniques based on planar lithography~\cite{Kirtley_2016}, focused ion beam patterning of a cantilever coated with a superconducting film~\cite{wyss_magnetic_2021}, and self-aligned deposition on a quartz pipette~\cite{finkler_self-aligned_2010}. The latter technique, used to produce a so-called SQUID-on-tip, combines the smallest SQUID sensors, down to \SI{39}{nm} in diameter, with the highest magnetic flux sensitivity~\cite{anahory_squid--tip_2020}. In addition, they have been demonstrated to operate under magnetic fields up to \SI{5}{\tesla}~\cite{bagani_sputtered_2019} and are excellent probes of local dissipation, due to the exquisite temperature sensitivity of the SQUID's Josephson junctions (JJs)~\cite{halbertal_nanoscale_2016}.  

These probes were first fabricated in 2010 by thermally evaporating Al on the apex of a pulled quartz capillary according to a three-step self-aligned deposition method pioneered by Finkler \textit{et al.}~\cite{finkler_self-aligned_2010}. In 2014, Vasyukov \textit{et al.} realized Pb devices with diameters of \SI{46}{nm} and flux noise of \SI{50}{n\Phi_0/\sqrt{\hertz}}~\cite{vasyukov_scanning_2013}. The fabrication process relies on directional vapor deposition techniques, such as thermal or electron beam evaporation. Because of fractionation, this type of evaporation is unsuitable for depositing alloys, limiting the resulting SQUID-on-tip probes to the elemental superconductors. Despite the excellent properties of such SQUID-on-tip probes, most consist of a thin superconducting film of, e.g., Pb, In, or Sn, which rapidly oxidizes under ambient conditions~\cite{anahory_squid--tip_2020}. This reactivity complicates the mounting, handling, and storage of these probes, making them difficult to use and limiting their widespread adoption. In addition, the high surface mobility of these materials requires cryogenic cooling of the pulled capillary during thermal evaporation in order to limit island formation. SQUID-on-tip probes made of ambient-stable Nb have been fabricated using electron beam evaporation by Vasyukov \textit{et al.}~\cite{vasyukov_scanning_2013} without cryogenic cooling of the capillary. In this process, an additional aluminum oxide buffer layer was deposited on the quartz capillary to prevent contamination of the Nb film. Although these probes are more robust than SQUID-on-tip probes of other elemental superconductors, the evaporated Nb devices exhibit significantly higher magnetic flux noise of \SI{3.6}{\micro\Phi_{0}/\sqrt{\hertz}}.

In 2019, Bagani \textit{et al.} demonstrated a SQUID-on-tip made of a superconducting alloy, MoRe, using a fabrication technique based on collimated magnetron sputtering~\cite{bagani_sputtered_2019}. Magnetron sputtering allows for the high-quality deposition of a wide range of superconducting materials beyond elemental superconductors, including alloys and multilayer structures. To impose directionality on the typically isotropic deposition of conventional magnetron sputtering, the sputtering source is enclosed in a small chamber equipped with a narrow slit. Ar gas is introduced into the small chamber, while the main chamber is kept in an ultrahigh-vacuum (UHV). Sputtered material, along with Ar gas, is forced through the narrow slit due to the pressure difference, resulting in collimated deposition. Although collimation is achieved, a large sputtering rate is required in the small chamber to achieve a useful flux of material from the slit. After just a few depositions, this large rate can lead to electrical shorting of the sputtering source, due to material flaking. As a result, regular opening and cleaning of the small chamber is required, which is detrimental to the achievable vacuum and, ultimately, the quality of the deposited film.

\begin{figure*}[t]
  \centering
  \includegraphics[width=0.98\textwidth]{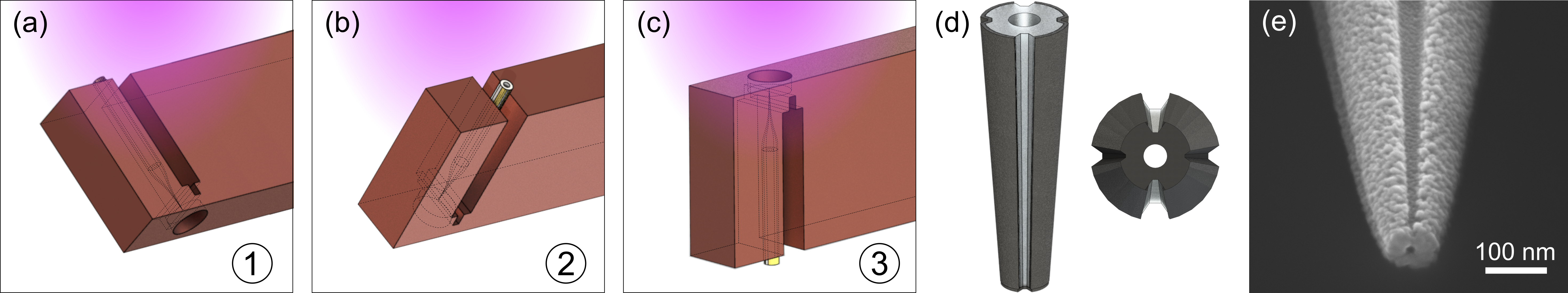}
  \caption{ Illustration of the SQUID-on-tip holder for magnetron sputtering deposition. The holder is rotated at deposition angles of (a)~$+\SI{120}{\degree}$, (b)~$-\SI{120}{\degree}$, and (c)~\SI{0}{\degree}, with respect to the direction of the sputtering source. The plasma is schematically shown in purple. (d)~Schematic drawing of a pulled four-groove capillary after film deposition with views from the side and bottom. (e) SEM micrograph of a Nb SQUID-on-tip with geometrical diameter of 55 nm.}
  \label{Fig1}
\end{figure*}

Here, we demonstrate the fabrication of SQUID-on-tip probes via a simplified magnetron sputtering process, which does not require an additional collimation chamber. Rather, directional deposition is achieved via an appropriately designed tip holder. This solution reduces the required opening of the sputtering chamber, substantially improves film quality, and allows for the exploration of superconducting materials for SQUID-on-tip fabrication. We demonstrate the process by fabricating Nb and MoGe SQUID-on-tip probes, which are less than \SI{100}{\nano\meter} in diameter, have magnetic flux noise down to \SI{300}{\nano\Phi_{0}/\sqrt{\hertz}}, and operate in fields as high as \SI{2.5}{\tesla}. Both the Nb and MoGe probes are stable in ambient conditions, allowing for easy handling and mounting.

We fabricate SQUID-on-tip probes by depositing a superconducting film onto pulled quartz capillaries. The capillaries, which have an outside diameter of \SI{1}{\milli\meter} and an inside diameter of \SI{0.4}{\milli\meter} (Friedrich \& Dimmock Inc.), are designed with four grooves, which are equally spaced around the external circumference. Once pulled by a laser-puller (Sutter Instrument P-2000/G), the outer diameter of the sharp tip can be less than \SI{100}{\nano\meter}. As illustrated in Fig.~\ref{Fig1}~(d), the grooves are preserved throughout the pulling process and help to define the pattern of the subsequently deposited superconducting film. Even though only two grooves are required for the two superconducting leads, four grooves help to preserve the capillary's circular geometry.

Prior to superconducting film deposition, capillaries are mounted in an electron-beam evaporator, in which a horizontal Ti/Au (2/\SI{5}{\nano\meter}) strip is deposited at a distance of \SI{350}{\micro\meter} from the apex of each tip using a shadow mask. In a finished SQUID-on-tip, this metallic strip works as a resistive shunt of 4--\SI{10}{\Omega} bridging the two superconducting electrodes, suppressing undesirable hysteresis, and protecting the device from electrostatic discharge during handling.

Next, each capillary is mounted in a dedicated UHV magnetron sputtering system with a \textit{in situ} rotatable holder and two commercial sputtering sources (AJA A320). The main chamber is pumped out and baked to reach a base pressure of approximately \SI{2d-9}{\milli\bar}. During deposition, a continuous flow of \SI{30}{sccm} Ar gas is supplied to the main chamber, maintaining a constant pressure of \SI{3.5d-3}{\milli\bar} with less than \SI{70}{\watt} of DC power required. In this work, the sputtering targets used are Nb (99.99\%), Mo$_{0.79}$Ge$_{0.21}$ (99.95\%), and Ti (99.95\%). 

The rotatable capillary holder is located approximately \SI{80}{\milli\meter} from the sputtering target surface and its shape, illustrated in Fig.~\ref{Fig1}~(a)-(c), has a considerable impact on the coating process. The plasma generated during the sputtering process, also known as an electrically conducting fluid, bends towards protruding conducting surfaces due to the locally deformed electric potential. The 1 to \SI{2}{\milli\meter} wide slots machined into the holder are designed to direct the sputtered material onto either one of the two sides or onto the tip of the capillary, depending on the holder's orientation with respect to the sputtering source. As illustrated in Fig.~\ref{Fig1}~(a)-(b), in the first two depositions of the process, the holder is oriented such that the capillary is tilted $\pm 120\grad$ from the direction of the sputtering source; in this way, one of its sides is exposed to the plasma represented in purple. These two depositions form the leads of the SQUID. The third deposition, carried out with the tip of the capillary pointing at and exposed to the sputtering source, is shown in Fig.~\ref{Fig1}~(c) and produces the SQUID loop at the apex of the capillary.

In addition to exploiting the directional deposition imposed by the capillary holder, the grooved quartz capillaries favor the formation of a well-defined gap between the two superconducting leads~\cite{bagani_sputtered_2019,anahory_squid--tip_2020}. Moreover, the grooves assist in the formation of the two narrow constrictions that form weak-link Josephson junctions -- and therefore the SQUID -- at the apex of the capillary. 

Using this process, we fabricate both Nb and MoGe SQUID-on-tip probes, which are realized by depositing 25--\SI{30}{\nano\meter} and 35--\SI{40}{\nano\meter} of material at a rate of \SI{3}{\angstrom/\second} in each of the three deposition steps, respectively. The Nb devices also include a \SI{3}{\nano\meter} base layer of Ti to avoid contamination from the quartz capillary, as well as a \SI{3}{\nano\meter} capping layer to protect from oxidation. A resulting Nb SQUID-on-tip probe is shown in a scanning electron micrograph (SEM) in Fig.~\ref{Fig1}~(e), with a geometrical diameter of \SI{55}{\nano\meter}.

\begin{figure*}[ht]
  \centering
  \includegraphics[width=0.75\textwidth]{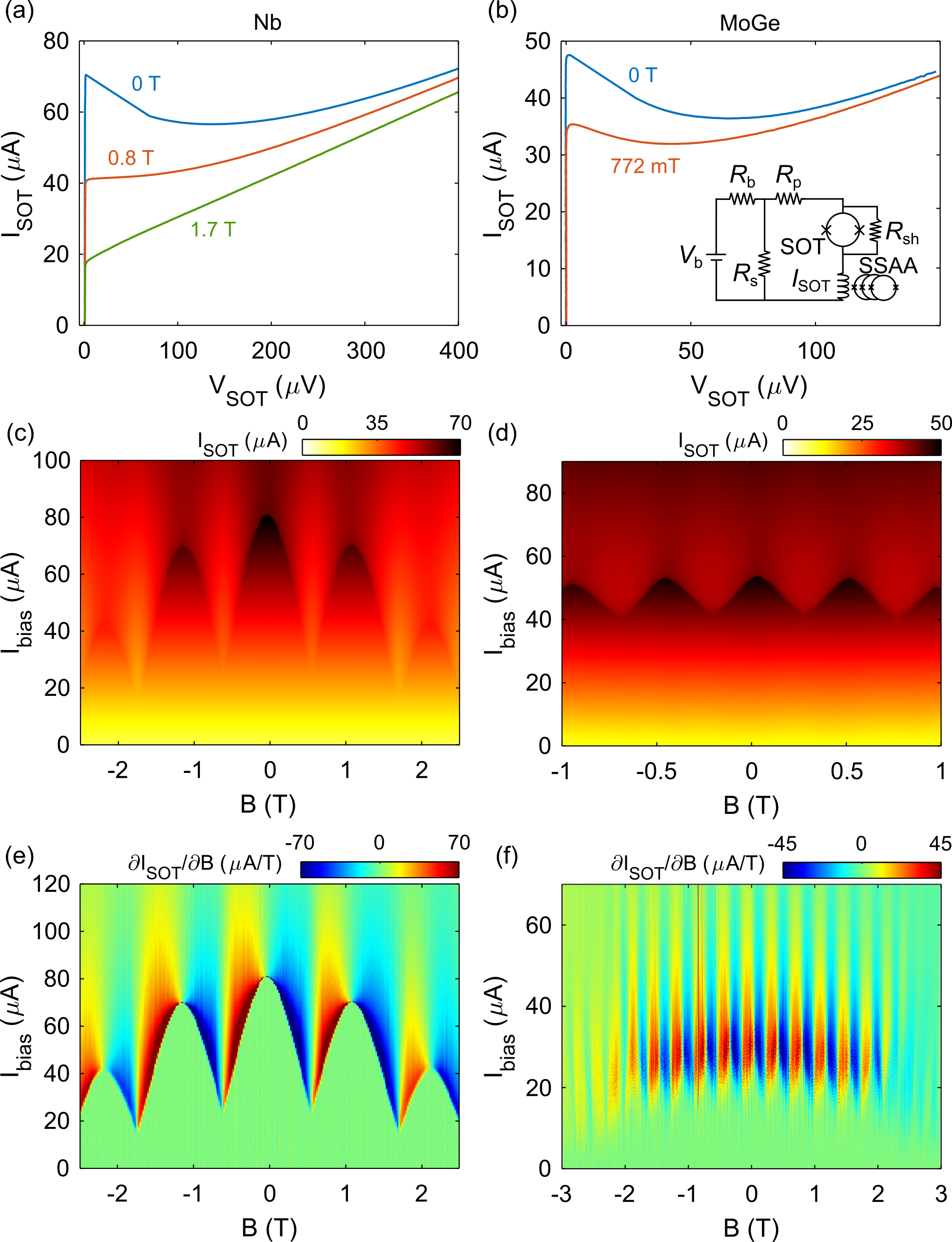}
  \caption{Current-voltage characteristics at \SI{4.2}{\kelvin}. Current through the SQUID-on-tip $I_\text{SOT}$ as a function of the voltage across it $V_\text{SOT}$ at different magnetic fields applied along the probe axis for (a)~Nb ($R_\text{p}=\SI{0.7}{\Omega}$, $R_\text{sh}=\SI{5}{\Omega}$) and (b)~MoGe  ($R_\text{p}=\SI{0.5}{\Omega}$, $R_\text{sh}=\SI{7}{\Omega}$) devices. The inset shows a simplified diagram of the electronic circuit used. Color-coded maps of $I_\text{SOT}$ versus bias current $I_\text{bias}$ and applied magnetic field $B$ for (c)~Nb and (d)~MoGe SQUID-on-tip showing a characteristic quantum interference pattern corresponding to an effective diameter of \SI{48}{\nano\meter} and \SI{74}{\nano\meter}, respectively. Corresponding color-coded maps of the magnetic response $\partial I_\text{SOT}/\partial B$ for the (e) \SI{48}{\nano\meter} Nb and (f) \SI{82}{\nano\meter} MoGe SQUID-on-tip probes.}
  \label{Fig2}
\end{figure*}

The current-voltage characteristics of Nb and MoGe SQUID-on-tip probes are measured at \SI{4.2}{\kelvin} and shown in Fig.~\ref{Fig2} at different magnetic fields applied along the probe axis. The electrical circuit used for characterization, illustrated in the inset of Fig.~\ref{Fig2}~(b), consists of a voltage source in series with a large bias resistor of \SI{6.1}{\kilo\Omega}, providing a bias current $I_{\text{bias}} = V_\text{b}/R_\text{b}$. A small shunt resistance $R_\text{s}=\SI{3}{\Omega} \ll R_\text{b}$ provides an effective bias voltage to the SQUID-on-tip. $R_\text{p}$ quantifies the parasitic resistance of the wires and contacts, while $R_\text{sh}$ is the resistance of the Ti/Au shunt bridging the superconducting electrodes close to the apex of the tip. The current through the SQUID-on-tip, $I_\text{SOT}$, is measured using a cryogenic SQUID series array amplifier (Magnicon), and the voltage across the SQUID-on-tip, $V_\text{SOT}$, is obtained from $V_\text{b}$ according to the electronic circuit.

In Fig.~\ref{Fig2}(a) and (b), plots of the current as a function of the voltage across the Nb and MoGe SQUID-on-tip probes show the amplitude of the critical current at different applied fields. The current increases until its critical point where the SQUID switches to a non-zero voltage state.
For non-hysteretic probe characteristics, i.e.\ overdamped behavior, the Stewart-McCumber parameter $\beta_\text{c}=2\pi I_\text{c}R^{2}C/\Phi_{0}$ has to be less than unity, where $I_\text{c}$ is the critical current, $R$ is the normal state resistance, $C$ the self-capacitance of the JJ, and $\Phi_{0}$ is the flux quantum. We estimate $\beta_\text{c}$ by determining the retrapping current, which we measure by decreasing the applied bias current from the normal state back to the superconducting state. Both probes show slightly overdamped behavior with $\beta_\text{c}^\text{Nb} \approx 0.93$ and $\beta_\text{c}^\text{MoGe} \approx 0.87$, with a critical current $I_\text{c}$ of \SI{70}{\micro\ampere} and \SI{48}{\micro\ampere} at zero applied field, respectively.

Fig.~\ref{Fig2}(c) and (d) show the quantum interference patterns of the Nb and MoGe SQUID-on-tip probes. From the magnetic-field periodicity of the modulation patterns, we calculate effective SQUID diameters of $d=2\sqrt{\Phi_{0}/\pi\Delta B} = \SI{48}{\nano\meter}$ and \SI{74}{\nano\meter}, for the Nb and MoGe devices, respectively. From these patterns, we also calculate the magnetic response $\partial I_\text{SOT}/\partial B$ of both SQUID-on-tip as a function of bias current $I_{\text{bias}}$ and applied magnetic field $B$, shown in Fig~\ref{Fig2} (e) and (f). The Nb interference pattern shows a sinusoidal shape and a large modulation depth of $I_\text{c}(B)$, resulting in a large magnetic response over a wide field range. The MoGe pattern is less ideal, showing a smaller modulation depth of $I_\text{c}(B)$ and consequently a smaller magnetic response, because of the high kinetic inductance of the MoGe film. For optimal SQUID characteristics, the screening parameter $\beta_\text{L}=2LI_\text{c}/\Phi_{0}$ must be close to unity, where $L$ is the inductance of the SQUID loop. Assuming symmetric JJs, we estimate from the modulation of $I_\text{c}$ the screening parameter $\beta_\text{L}$ for Nb and MoGe SQUID-on-tips: $\beta_\text{L}^\text{Nb} \approx 0.66$ and $\beta_\text{L}^\text{MoGe} \approx 2.75$. The SQUID loop inductance $L=L_\text{g}+L_\text{K}$ comprises the geometrical inductance $L_\text{g}$ of the SQUID loop and the kinetic inductance $L_\text{K}$. Due to the small loop size, the geometrical contribution can be neglected, while the kinetic contribution depends mostly on the material properties and film thickness. Improving the magnetic response of the MoGe SQUID-on-tip probe would require a superior film quality, potentially realized by annealing the device, which can reduce its kinetic inductance and therefore increase the modulation depth of $I_\text{c}(B)$.

The high magnetic response even at large applied magnetic fields of ~\SI{2}{\tesla} distinguishes these probes from planar scanning SQUID probes~\cite{LamNanoscaleSQUIDOperating2011}, making them particularly useful for measuring high-field condensed matter phenomena.  The geometry of the SQUID-on-tip allows operation in large applied fields, due to the mostly in-plane field experienced by the superconducting leads~\cite{chenOnchipSQUIDhighfields2010}.
The Nb SQUID-on-tip presents an extended operation range of \SI{2.5}{\tesla} in comparison to previously reported Nb SQUID-on-tip of \SI{1}{\tesla} range~\cite{vasyukov_scanning_2013}. The thin Ti base and capping layers likely lead to the formation of a NbTi alloy, which contributes to a larger H$_\text{c2}$ than expected from a pure Nb thin film. MoGe SQUID-on-tip shows persistent oscillations with a good magnetic response up to \SI{2}{\tesla} and reduced response from there up to \SI{3}{\tesla}. The asymmetry in the magnetic response in positive and negative applied fields is due to two weak-link Josephson junctions of slightly different sizes. The MoGe SQUID-on-tip presents a higher magnetic response than the previously reported annealed MoRe SQUID-on-tip of similar loop size~\cite{bagani_sputtered_2019}, however, with reduced applied field operation range.

\begin{figure}[h]
  \centering
  \includegraphics[width=0.47\textwidth]{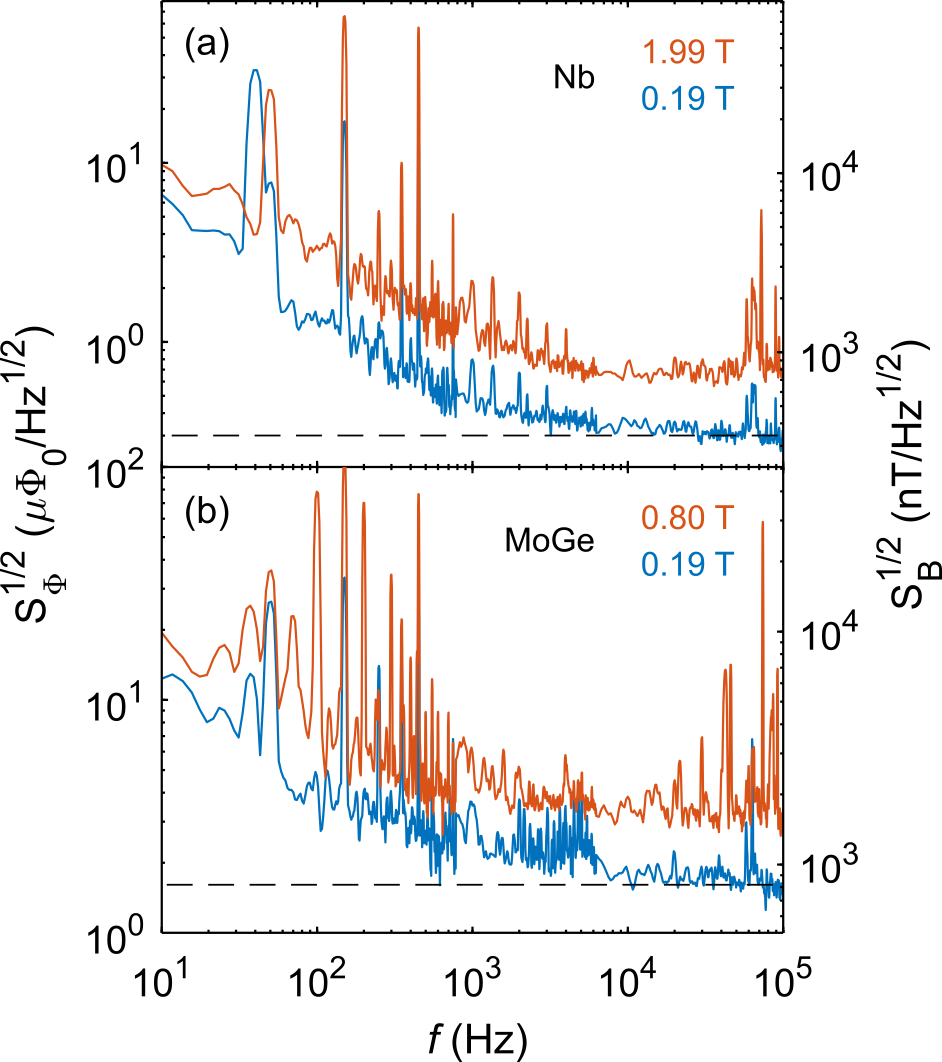}
  \caption{ Spectral density of the magnetic flux noise $S_{\Phi}^{1/2}$ and magnetic field noise $S_\text{B}^{1/2}$ measured in low and high applied magnetic fields at \SI{4.2}{\kelvin} for a (a) \SI{48}{\nano\meter} Nb SQUID-on-tip, and (b) \SI{74}{\nano\meter} MoGe SQUID-on-tip.}
  \label{Fig3}
\end{figure}

The magnetic flux and field spectral densities of a Nb SQUID-on-tip with effective diameter of \SI{48}{\nano\meter} are shown in Fig.~\ref{Fig3}(a). They are determined at fixed $V_\text{b}$ and at the indicated magnetic fields by measuring the current noise of the probe and taking into account the magnetic response shown in Fig.~\ref{Fig2}(e), as well as the effective SQUID loop area for the flux noise. At lower frequencies the spectrum is dominated by 1/f-noise. In the white noise regime above \SI{10}{\kilo\hertz}, the magnetic field noise reaches a minimum value of $S_\text{B}^{1/2}=$~\SI{340}{\nano\tesla/\sqrt{\hertz}}, which corresponds to a magnetic flux noise of $S_{\Phi}^{1/2}=S_\text{B}^{1/2}(\pi r^{2})=$~\SI{300}{\nano\Phi_{0}/\sqrt{\hertz}} in low fields. This value is one order of magnitude better than the one previously reported for a Nb SQUID-on-tip~\cite{vasyukov_scanning_2013}, likely due to the improved quality of the sputtered versus the evaporated film. At high applied magnetic fields, there is a slight decrease in performance with $S_{\Phi}^{1/2}=$ \SI{600}{\nano\Phi_{0}/\sqrt{\hertz}}. For some nanometer-scale imaging applications, such as the mapping of defect spins or of weakly magnetic samples, it is important to quantify the spin sensitivity of the probes. In the case of a spin in the center of the SQUID loop oriented perpendicular to it, the spin noise is estimated by $S_\text{n}^{1/2}=S_{\Phi}^{1/2}r/r_\text{e}$, where $r_\text{e}$ is the classical electronic radius~\cite{Ketchen_1989}. From the spectral density of the spin noise, we obtain the limit for spin sensitivity: for the Nb SQUID-on-tip, a base value of $S_\text{n}^{1/2}=$~\SI{2.7}{\mu_{B}/\sqrt{\hertz}} is measured in low fields and \SI{6.6}{\mu_{B}/\sqrt{\hertz}} in higher fields.

The spectral density of the \SI{74}{\nano\meter} MoGe SQUID-on-tip in Fig~\ref{Fig3}(b) shows a magnetic field noise of $S_\text{B}^{1/2}=$~\SI{800}{\nano\tesla/\sqrt{\hertz}}, corresponding to a magnetic flux noise of $S_{\Phi}^{1/2}=$~\SI{1.6}{\micro\Phi_{0}/\sqrt{\hertz}} in low applied field, and \SI{3}{\micro\Phi_{0}/\sqrt{\hertz}} in higher fields. The spin noise for MoGe is found to be from $S_\text{n}^{1/2}=$ 20 to \SI{35}{\mu_{B}/\sqrt{\hertz}}, which is comparable to the values reported for the annealed MoRe SQUID-on-tip~\cite{bagani_sputtered_2019}. 

We demonstrate a simplified fabrication of SQUID-on-tip probes made of elemental and composite materials, including multilayered structures, using magnetron sputtering deposition. The technique, which relies on a specially designed deposition holder and grooved capillaries, enabled the fabrication of a Nb SQUID-on-tip with sub-100-nm loop size, improved magnetic flux sensitivity of \SI{300}{\nano\Phi_{0}/\sqrt{\hertz}}, and critical field up to over \SI{2.5}{\tesla} ($H_\text{c2}$), likely due to the Ti base and cap layers. The fabricated devices are robust to ambient conditions, which make them easy to handle and mount. The technique should ease the fabrication of SQUID-on-tip devices from a variety of superconducting alloys and multilayers extending the critical temperature and field capabilities of these probes and facilitating their application to nanometer-scale magnetic and thermal imaging.

\bigskip
We thank Sascha Martin and his team in the machine workshop of the
Physics Department at the University of Basel.  We acknowledge the
support of the Canton Aargau; the Swiss National Science
Foundation via Project Grant No. 200020-159893 and Sinergia Grant
Nanoskyrmionics (Grant No. CRSII5-171003); the European Commission under H2020 FETOpengrant "FIBsuperProbes" (Grant No. 892427); and the University of Basel via Universit\"{a}t Basel Forschungsfonds (Project No. 4637580).

\bigskip
The authors have no conflicts to disclose.

\bigskip
The data that support the findings of this study are available from
the corresponding author upon reasonable request.


%

\end{document}